\documentclass[preprint,12pt]{elsarticle}
\usepackage{graphicx,amssymb}

\journal{Journal of Computational and Theoretical Nanoscience}

\def\p{\mathrm{p}}

\def\b{\mathrm{b}}
\def\d{\mathrm{d}}
\def\v{\mathrm{v}}
\def\m{\mathrm{m}}
\def\mp{{\mathrm{m}^\prime}}

\def\vr{\vec{r}}
\def\vn{\vec{n}}
\def\vv{\vec{v}}
\def\ve{\vec{E}}
\def\vf{\vec{F}}
\def\va{\vec{A}}
\def\vb{\vec{B}}
\def\vp{\vec{p}}
\def\vnabla{\vec{\nabla}}

\def\cf{{\cal F}}

\begin{document}

\begin{frontmatter}

\title{Modeling comminution processes in ball mills as a canonical ensemble}

\author{G. K. Sunnardianto$^\mathrm{a}$, Muhandis$^\mathrm{a}$, F. N.
Diana$^\mathrm{b}$, L.T. Handoko$^\mathrm{a,b}$\footnote{Corresponding
author.\\E-mail address : laksana.tri.handoko@lipi.go.id (L.T. Handoko)}}

\address{$^\mathrm{a}$Group for Theoretical and Computational Physics, Research Center for Physics, Indonesian Institute of Sciences (LIPI), Kompleks Puspiptek Serpong, Tangerang 15310, Indonesia\\
$^\mathrm{b}$Department of Physics, University of Indonesia, Kampus UI Depok, Depok 16424, Indonesia}

\begin{abstract}
A new approach to describe comminution processes in general ball mills as a
macroscopic canonical ensemble is proposed. Using  hamiltonian method, the 
model is able to take simultaneously into account the internal dynamics from
 mechanical motions inside the vial and external effects like
electromagnetic and gravitational forces. Relevant physical observables are
extracted using statistical mechanics approach through partition function at
finite temperature. The method enables numerical calculation using Monte Carlo
technique to obtain, for instance particle number evolution in term of system
temperature. It is argued that the method is experimentally more verifiable than
the conventional approaches based on geometrical displacements. An example of
simulation for typical spex mill is also given.
\end{abstract}


\end{frontmatter}

\section{Introduction}
\label{intro}

The comminution processes in recent years attract the attention among scientists and engineers due to the increasing demand of ultrafine powders for nanotechnology applications in many areas. The demand then requires the improvement of comminution equipments like ball mills, roller mills and so on. Unfortunately, the development of such comminution equipments always contains a lot of uncertainties due to a wide range of unknown parameters. These, in fact, lead to significant statistical errors. In order to overcome such problems, several models have been developed to quantitatively describe comminution process in various types of mills \cite{mishra92,mishra94,mishra95,mishra01,poschel}.

On the other hand, mathematical modeling and simulation may provide prior
information and constraint to the unknown parameter ranges which should be
useful to develop more optimized experimental strategy in comminution processes.
However, in most cases of mathematical models, the physical observables like
grain-size etc are extracted from a set of equation of motions (EOMs). Such
EOM's are considered to govern as complete as possible the dynamics of the
system, from the mechanical motions to the evolution of grain-size distribution.
This approach is obviously suffered from the nonlinearities of the equations
under consideration, and then the requirement of high computational power to
solve them numerically. This fact often discourages a quantitative and
deterministic approach for the simulation of such system. These nonlinear
effects like chaotic behavior of the sphere motions within the mill encourages
some works modeling the system using semi-empirical approaches \cite{manai}.
However most of semi-empirical models require a large number of experimental
data based on prior observations \cite{davis}, or measured variables obtained
from simulation results by other authors \cite{maurice1990,maurice1996}.

More empirical approach is based on the physically realistic modelization of the ball mill system \cite{delogu,wang}. This approach in general deals with three aspects : (1) evaluation of milling bodies dynamics and energetic inputs transferred to powders; (2) description of the effects of such inputs on powders breakage; (3) description of powders evolution in terms of particle size distribution \cite{concas}. In a recent work \cite{concas}, a comprehensive study on this line for the case of spex mixer / mill was  performed by deploying the 3D simulation for milling bodies motion, and the population balance method to describe the granulometric evolution. Then, both are related through the energetic inputs in the population balance formalism which is obtained from the simulation of milling bodies motion.

In this paper we propose a novel model and approach combining the deterministic
approach for milling bodies motion, and the statistical approach to relate them
with considerable macroscopic physical observables. This work is devoted to
overcome the following problems in some conventional approaches :
\begin{itemize}
\item Experimentally it is almost impossible to trace the geometrical displacements of all  matters in a vial with proper time resolution to verify the models which are based on the classical EOM. This problem is getting worse as one simulates a system with matters at few nanometers scale with comparable size of time-space resolution.
\item Solving a set of EOMs numerically, and then performing a simulation with high accuracy (enough time resolution) require huge efforts on both computing capacity and running time.
\item Taking into account the external circumstances around the vial like electromagnetic field and so forth. This might be interesting when one considers a comminution process which can reach the level of few nanometers.
\end{itemize}
Therefore this work is intended to provide a tool for a nanometer system in a
vial by developing direct relations between the vial internal dynamics with some
external physical observables which should be more measurable. We should remark
here that the vial internal dynamics is yet described empirically using physical
modelization approach.

Further, rather solving the EOM's governing the whole dynamics, we use the
hamiltonian approach to model all relevant interactions, and extract the
physical observables through partition function by treating the system as a
canonical ensemble at finite temperature. As a consequence, instead of
observing the geometrical evolution of matters in term of time in a ball mill,
we can investigate the particle number distribution in term of temperature. This
introduces a novel method relating the internal dynamics with the macroscopic
physical parameter like temperature, rather than time and geometrical
displacements which are in most cases difficult to realize. The preliminary
work but with incomplete hamiltonian on this matter has been reported in our
previous work \cite{handoko}, however
the present paper comprises more complete theoretical formalism
including the external forces and the real simulation using Monte Carlo
integration method.

The paper is organized as follows. First, after this introduction we present 
the model and explain the underlying knowledge in detail. Before summarizing
the results, numerical analysis and simulation for the case of ball mill with a
structure similar to the  well-known spex mixer / mill are discussed.

\section{The model}

The whole system is modeled empirically using hamiltonian method. First we construct the total hamiltonian describing the dynamics in the ball mills. It is further followed by formulating the partition function and extracting the relevant thermodynamics observables.

\subsection{The dynamics}

In our model, the dynamics of each 'matter' in the system, i.e. balls and
powders inside the vial, are described by a hamiltonian $H_\m(\vr,t)$. The index
$\m$ denotes the powder ($\p$) or ball ($\b$) and $\vr = (x,y,z)$. The
hamiltonian contains some terms representing all relevant interactions working
on the matters inside the system as follow, 
\begin{equation}
 H_\m = H_0 + V_{\m-\m} + V_{\m-\v} + V_{\m-\m^\prime} + V_\mathrm{ext} \; ,
\label{eq:h}
\end{equation}
with $\v$ denotes the vial, while $H_0$ is the free matter hamiltonian
containing the kinetic term,
\begin{equation}
 H_0 = \frac{1}{2 m_\m} \sum_{i=1}^{n_\m} \left| \left( \vp_\m \right)_i \right|^2 \; ,
\label{eq:h0}
\end{equation}
where  $n_\m$ is the matter number, $m_\m$ and $\vp_\m$ are the matter mass and
momentum respectively. Throughout the paper we assume that the mass or size
evolution of matters is uniform for the same matters.

The matter self-interaction $V_{\m-\m}$, the matter--vial interaction
$V_{\m-\v}$ and the interactions between different matters may be induced by,
for instance,  impact ($V^{\mathrm{imp}}$) and Coulomb ($V^{\mathrm{Coul}}$)
potentials,
\begin{eqnarray}
 V^{\mathrm{imp}}_{\m-\mp}(\vr,t) & = & -\sum_{i=1}^{n_\m} \sum_{j=1}^{n_\mp}  \int_0^{\left(\xi_{\m\mp}\right)_{ij}} \d \left(\xi_{\m\mp}\right)_{ij} \, \vn \cdot \left( \vf^{\mathrm{imp}}_{\m\mp} \right)_{ij} \; , 
 \label{eq:vimp}\\
 V^{\mathrm{Coul}}_{\m-\mp}(\vr) & = & Q_\m Q_\mp \, \sum_{i=1}^{n_\m}\sum_{j=1}^{n_\mp} \frac{1}{\left| \left( \vr_\m \right)_i - \left( \vr_\mp \right)_j \right|} \; ,
\label{eq:vcoul}
\end{eqnarray}
with $Q_\m$ is the matter charge, while $\m,\mp : \v, \p, \b$ and $\vn$ is the
unit normal vector. These potentials are considered describing the mechanical
and static electrical properties of the matters. The first potentials should in
fact represent the whole classical dynamics among the matters, i.e. the impact
forces among balls and powders. This form will be clarified soon below. The
Coulomb potential disappears if the interacting matters have neutral charges.
Also it works only in a short range of distance. Therefore, it should be
negligible for some physics at nanometers scale as considered in the present
case. The impact forces between the vial surface and balls or powders are 
treated in the same way using Eq. (\ref{eq:vimp}) by considering that the
surface is formed by a set of fixed spheres \cite{concas}. This is inline
with the simulation in the last section where the space displacement in the
vial is at a distance  comparable with the desired powder size, i.e. few
tens nanometers at the largest.

On the other hand, instead of Eq. (\ref{eq:vimp}) we can consider a simpler
'effective potential' like  the  harmonic oscillator
$V^{\mathrm{osc}}_{\m-\mp}(\vr) = \frac{1}{2} \, k_{\m\mp} \, \Delta \vr^2$ to
represent the whole mechanical properties in terms of 'effective coupling'
$k_{\m\mp}$. In this case, if $m_\m \gg m_\mp$, the potential can be rewritten
as $V_{\m-\mp} = \frac{1}{2} \, m_\m \, {\omega_\m}^2 \, \Delta \vr^2$ since
$\omega_\m \equiv \sqrt{{k_{\m\mp}}/{m_\m}}$. Actually this is the case of
ball--powder interactions since $m_\b \gg m_\p$ by the order of namely
$O(10^2)$. Nevertheless, in contrast  with its simplicity, $V^{\mathrm{osc}}$
absorbs the time dependencies and another interesting physical parameters
characterizing the material properties like viscoelasticity, Young modulus etc.
The time dependency is important to directly relate the system temperature with
physical observables through finite temperature partition function as shown in
the next subsection. Therefore, in this paper we take the impact potential to
represent the mechanical properties in the system.

The impact potential in Eq. (\ref{eq:vimp}) is induced by the impact force consists of normal and tangential components \cite{concas}, $\vf_{\m\mp}^{\mathrm{imp}}(\vr,t) = \vf_{\m\mp}^N(\vr,t) + \vf_{\m\mp}^T(\vr,t)$. The normal component may be written \cite{brilliantov},
\begin{equation}
 \vf_{\m\mp}^N(\vr,t) = \left [ \frac{2 \Upsilon_{\m\mp}}{3 (1 - v_{\m\mp}^2)} \sqrt{R_{\m\mp}^\mathrm{eff}} \left ( \xi_{\m\mp}^{{3}/{2}} + \frac{3}{2} A_{\m\mp} \sqrt{\xi_{\m\mp}} \, \frac{\d \xi_{\m\mp}}{\d t}\right) \right] \vn  \; .
\label{eq:fn}
\end{equation}
Here the first term is the elastic part based on the Hertz contact law, and the second term is the dissipative one that takes into account material viscosity. $\Upsilon_{\m\mp}$ is the Young modulus and $v_{\m\mp}$ represents the Poisson ratio of the sphere material. The term $R_{\m\mp}^\mathrm{eff} = {(R_\m R_\mp)}/{(R_\m + R_\mp)}$ represents the effective radius, while $\xi_{\m\mp} = R_\m + R_\mp - | \vr_\m - \vr_\mp|$ is the displacement with $R_\m$ is the radius of interacting matter. $A$ is a dissipative parameter \cite{brilliantov,landau,hertzsch}, 
\begin{equation}
A_{\m\mp} = \frac{1}{3} \frac{{3 \eta_\mp - \eta_\m}^2}{3 \eta_\mp + 2 \eta_\m} 
	\left[ \frac{(1 - v_{\m\mp}^2 )( 1 - 2 v_{\m\mp})}{ \Upsilon_{\m\mp} \, v_{\m\mp}^2} \right] \; .
\label{eq:a}
\end {equation}
The viscous constants $\eta_\m$ and $\eta_\mp$ relate the dissipative stress tensor to the deformation tensor \cite{brilliantov,landau}.

There are several proposed formulations for the tangential component
$\vf_{\m\mp}^T (\vr,t)$. However it always assumes that the material slide upon
each other in the case where the Coulomb condition $\mu \left| \vf_{\m\mp}^N
\right| \leq \left| \vf_{\m\mp}^T \right|$ holds, otherwise some viscous
frictions occur \cite{saluena}. In particular it follows that
$\vf_{\m\mp}^T(\vr,t) \propto m_{\m\mp}^\mathrm{eff}$, where the effective mass
is $m_{\m\mp}^\mathrm{eff} \equiv {m_\m m_\mp}/{(m_\m + m_\mp)}$ \cite{concas}.
Obviously, in our case with large mass discrepancy between powder and ball,
i.e. ${m_\p}/{m_\b} \sim 0$ which leads to $m_{\p\b}^\mathrm{eff} \sim m_\p$,
the tangential impact force may be neglected for a good approximation. So, let
us from now consider the normal component dominated impact force, that is
$\vf_{\m\mp}^{\mathrm{imp}}(\vr,t) \sim \vf_{\m\mp}^N(\vr,t)$. This result
simply yields the impact potential as written in Eq. (\ref{eq:vimp}) due to the
Euler-Lagrange equation,
\begin{equation}
 \vf = -\frac{\d V}{\d \vr} + \frac{\d}{\d t} \left( \frac{\d V}{\d \vv} \right) \; ,
\label{eq:ele}
\end{equation}
since the dependency on matter velocity appears only in the tangential component $\vf_{\m\mp}^T(\vr,t)$ \cite{concas}.

Beside the interactions among the matters itself, it is also possible to take
into account the external potentials working on the whole system. For instance 
concerning the ball dynamics with relatively large ball size, the gravitational
potential,
\begin{equation}
 V^{\mathrm{grav}}_\mathrm{ext} = m_\m \, G \, \sum_{i=1}^{n_\m} \left( z_\m \right)_i \; ,
\end{equation}
might be important. Here, $G$ is the gravitational constant. On the other hand,
this should be less important for the powder dynamics due to its tiny size.

On the other hand, we may also incorporate the effect of external
electromagnetic field surrounding the system to the charged matters. The
potential is induced by the Lorentz force, $\vf_\m^{\mathrm{EM}} = Q_\m \, (\ve
+ \vv_\m \times \vb)$, which leads to,
\begin{equation}
 V_{\mathrm{ext}}^{\mathrm{EM}} = Q_\m \sum_{i=1}^{n_\m} \left[ \phi - \left( \vv_\m \right)_i \cdot \va \right] \; ,
\label{eq:vem}
\end{equation}
and satisfies Eq. (\ref{eq:ele}). $\phi$ and $\va$ are the electromagnetic scalar and vector potentials related to the electric and magnetic fields by $\ve = - \vnabla \phi - {\partial \va}/{\partial t}$ and $\vb = \vnabla \times \va$.
The inclusion of electromagnetic potential shifts the kinetic term in Eq. (\ref{eq:h0}) as follow,
\begin{equation}
 H_0 \longrightarrow H_{0 + \mathrm{EM}} = \frac{1}{2 m_\m} \sum_{i=1}^{n_\m} \left| \left( \vp_\m \right)_i - Q_\m \, \va \right|^2 + n_\m \, Q_\m \, \phi \; ,
\label{eq:h0em}
\end{equation}

From now, let us focus only on the dynamics of powders which is our main
interest in the sense of comminution process. From Eqs. (\ref{eq:h}),
(\ref{eq:h0}), (\ref{eq:vimp}), (\ref{eq:vcoul}) and (\ref{eq:vem}), the total
hamiltonian for the powder in our model is,
\begin{eqnarray}
 H_\p & = & \frac{1}{2 m_\p} \sum_{i=1}^{n_\p} \left| \left( \vp_\p \right)_i - Q_\p \, \va \right|^2 + n_\p \,Q_\p \, \phi  
	\nonumber \\
	&& - \frac{1}{2} \sum_{i(\neq j)=1}^{n_\p} \sum_{j=1}^{n_\p}  \int_0^{\left(\xi_{\p\p}\right)_{ij}} \d \left(\xi_{\p\p}\right)_{ij} \, \vn \cdot \left( \vf^{\mathrm{imp}}_{\p\p} \right)_{ij} 
	\nonumber \\
	&& - \sum_{m:\b,\v}\sum_{i=1}^{n_\p} \sum_{j=1}^{n_\m}  \int_0^{\left(\xi_{\p\m}\right)_{ij}} \d \left(\xi_{\p\m}\right)_{ij} \, \vn \cdot \left( \vf^{\mathrm{imp}}_{\p\m} \right)_{ij} \; ,
\label{eq:hp}
\end{eqnarray}
for $Q_\p \neq 0$. The last two potentials represent the total impact potential among powders; powders and vial; powders and balls respectively. Obviously we do not need to take into account the ball self-interaction $V^{\mathrm{imp}}_{\b-\b}$ nor ball-vial interaction $V^{\mathrm{imp}}_{\b-\v}$. This is actually the advantage of using hamiltonian method.

\subsection{Physical observables}

As mentioned briefly in introduction, the greatest advantage of deploying the
hamiltonian method is one can extract some physical observables without solving
the EOM's governing the system. Instead, in a canonical ensemble of matter
m one can consider the partition function,
\begin{equation}
 Z_\m = \int \prod_{i=1}^{n_\m} \d \vp_i \, \d \vr_i \; \mathrm{exp} \left[ 
 -\int_0^\beta \d t \, H_\m
 \right] \; ,
\label{eq:z}
\end{equation}
governed by a particular hamiltonian $H_\m$. Here, $\beta \equiv 1/{(k_B T)}$
with $k_B$ and $T$ are the Boltzman constant and absolute temperature. Having
partition function at hand, we can obtain some thermodynamics quantities  in the
system through relations namely,
\begin{equation}
 F_\m = -\frac{1}{\beta} \; \ln Z_\m \; ,
\label{eq:f}
\end{equation}
for free energy and,
\begin{equation}
 P_\m = -\frac{\partial F_\m}{\partial V} = -\frac{F_\m}{V} \; ,
\label{eq:p}
\end{equation}
for pressure in a vial with volume $V$.

In order to see the contributions from the interactions, it is more convenient to consider the normalized partition function,
\begin{equation}
 Z'_\m \equiv \frac{Z_\m}{Z_{0_\m}} = \frac{\displaystyle \int
\prod_{i=1}^{n_\m} \d \vp_i \, \d \vr_i \; \mathrm{exp} \left[ 
 -\int_0^\beta \d t \, H_\m
 \right]}{\displaystyle \int \prod_{i=1}^{n_\m} \d \vp_i \; \mathrm{exp} \left[
 -\int_0^\beta \d t \, H_{0_\m}
 \right]} \; ,
\label{eq:nz}
\end{equation}
and further,
\begin{equation}
 P'_\m = F'_\m \equiv \frac{F_\m}{F_{0_\m}} = \frac{\ln Z_\m}{\ln {Z_0}_\m}
\; .
\label{eq:nf}
\end{equation}

Performing the integral over time ($t$), we immediately obtain a temperature dependent partition function, while the integrals over $\vp_i$ are the decoupled gaussian integral which can be easily calculated. In the case of Eq. (\ref{eq:hp}) it gives,
\begin{equation}
  Z_\p = \left( \frac{2 \, m_\p \pi}{\beta} \right)^{{n_\p}/2} 
  \int \prod_{i=1}^{n_\p} \d \vr_i  \; \mathrm{exp} \left[ 
 -\int_0^\beta \d t \, H^\prime_\p
 \right] \; ,
\label{eq:z2}
\end{equation}
and, 
\begin{equation}
 Z'_\p = \int \prod_{i=1}^{n_\p} \d \vr_i  \; \mathrm{exp} \left[ 
 -\int_0^\beta \d t \, H^\prime_\p
 \right] \; ,
\label{eq:zp}
\end{equation}
where the interaction hamiltonian is,
\begin{eqnarray}
 H_\p^\prime & = & n_\p \, Q_\p \, \phi  
	- \frac{1}{2} \sum_{i(\neq j)=1}^{n_\p} \sum_{j=1}^{n_\p}  \int_0^{\left(\xi_{\p\p}\right)_{ij}} \d \left(\xi_{\p\p}\right)_{ij} \, \vn \cdot \left( \vf^{\mathrm{imp}}_{\p\p} \right)_{ij} 
	\nonumber \\
	&& - \sum_{\m:\b,\v} \sum_{i=1}^{n_\p} \sum_{j=1}^{n_\m}  \int_0^{\left(\xi_{\p\m}\right)_{ij}} \d \left(\xi_{\p\m}\right)_{ij} \, \vn \cdot \left( \vf^{\mathrm{imp}}_{\p\m} \right)_{ij} 
	 \; .
\label{eq:hpp}
\end{eqnarray}
Obviously, only the scalar potential of external electromagnetic field contributes to the  total energy of system under consideration. In other words, we can conclude here that in our model the magnetic field $\vb$ does not influence the ball mill system, but the electric field $\ve$ does.

\begin{figure}[t]
 \centering
 \includegraphics[width=10cm]{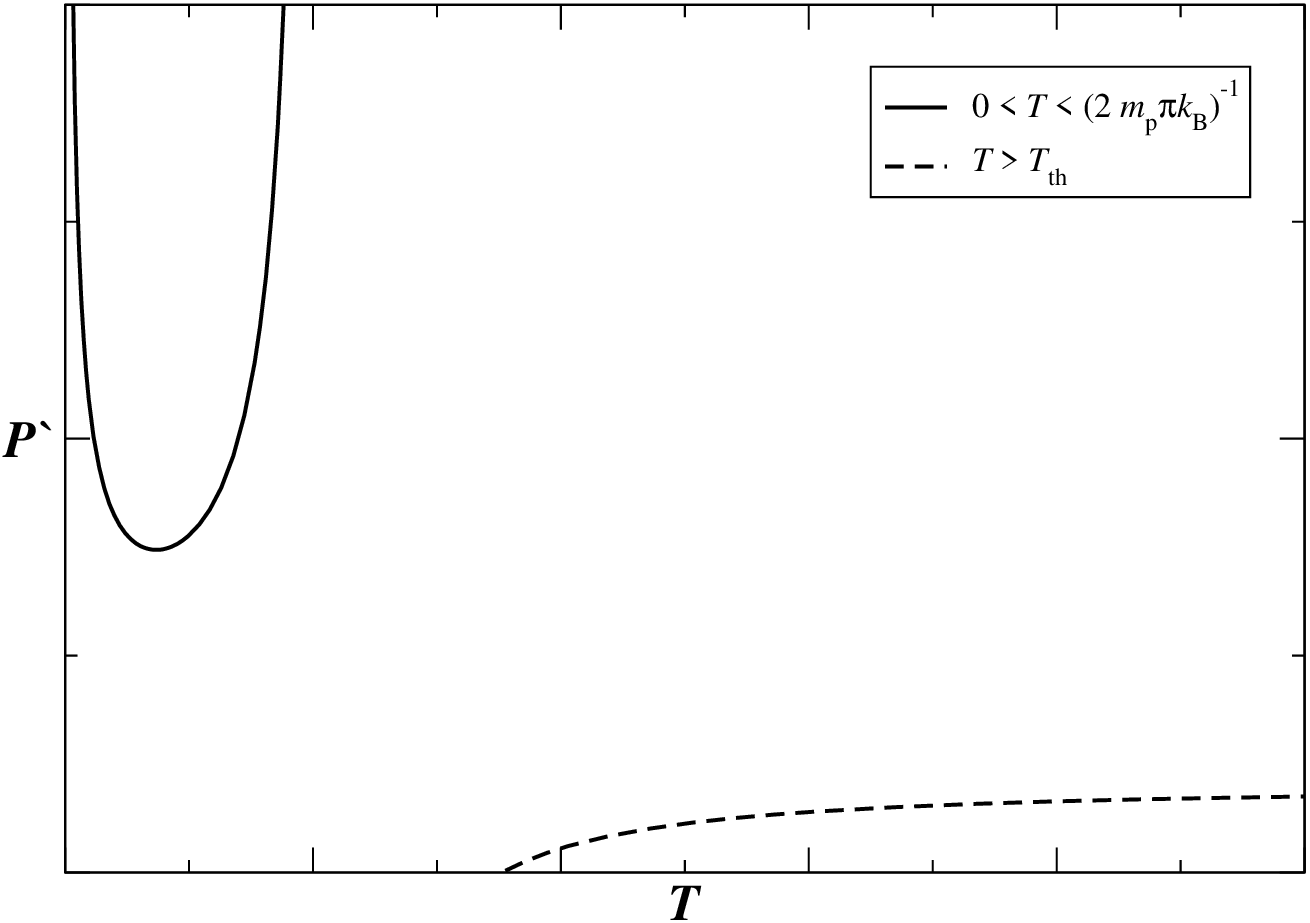}
 \caption{General contour of the normalized pressure as a function of system
temperature.}
 \label{fig:p}
\end{figure}

Moreover, we can perform the integration over time ($t$) and $\xi_{ij}$ to
obtain further,
\begin{eqnarray}
 Z'_\p & = & \int \prod_{i=1}^{n_\p} \d \vr_i  \; \mathrm{exp} \left[ -\beta
\left( 
 Q_\p \, \phi  
 - \frac{2}{15} \sum_{i(\neq j)=1}^{n_\p} \sum_{j=1}^{n_\p} \frac{\Upsilon_{\p\p}}{1 - v_{\p\p}^2} \sqrt{R_{\p\p}^\mathrm{eff}} \, \left(\xi_{\p\p}\right)_{ij}^{{5}/{2}} 
  \right.\right. \nonumber \\
  && \left.\left. - \frac{4}{15} \sum_{\m:\b,\v} \sum_{i=1}^{n_\p} \sum_{j=1}^{n_\m} \frac{\Upsilon_{\p\m}}{1 - v_{\p\m}^2} \sqrt{R_{\p\m}^\mathrm{eff}} \, \left(\xi_{\p\m}\right)_{ij}^{{5}/{2}} 
  \right) \right] \; .
\label{eq:zpf}
\end{eqnarray}
From this result, the thermodynamics observables are clearly not affected with the dissipative term, i.e. the second term in Eq. (\ref{eq:fn}). After performing same integration we obtain,  \begin{equation}
  P'_\p = 1 - \beta \, \cf \,  \ln^{-1} \left( \frac{2 \, m_\p \pi}{\beta}
\right) \; ,
\label{eq:pf}
\end{equation}
respectively with,
\begin{eqnarray}
 \cf & \equiv & 2 \,  \int \prod_{i=1}^{n_\p} \d \vr_i  \left[ 
  Q_\p \, \phi  
 - \frac{2}{15 \, n_\p} \sum_{i(\neq j)=1}^{n_\p} \sum_{j=1}^{n_\p} \frac{\Upsilon_{\p\p}}{1 - v_{\p\p}^2} \sqrt{R_{\p\p}^\mathrm{eff}} \, \left(\xi_{\p\p}\right)_{ij}^{{5}/{2}} 
  \right. \nonumber \\
  && \left. - \frac{4}{15 \, n_\p} \sum_{\m:\b,\v} \sum_{i=1}^{n_\p} \sum_{j=1}^{n_\m} \frac{\Upsilon_{\p\m}}{1 - v_{\p\m}^2} \sqrt{R_{\p\m}^\mathrm{eff}} \, \left(\xi_{\p\m}\right)_{ij}^{{5}/{2}} 
  \right] \; .
\label{eq:faux}
\end{eqnarray}
Eq. (\ref{eq:pf}) provides a general behavior for temperature-dependent pressure in the model, while the geometrical structure and motion of vial is absorbed in the function $\cf$. From Eq. (\ref{eq:pf}) clearly the physically meaningful regions are for $0 < T < (2 \, m_\p \, \pi \, k_B)^{-1}$ and $T \geq T_\mathrm{th}$. The later is equivalent to the condition,
\begin{equation}
 \cf \leq k_B \, T_\mathrm{th} \, \ln \left( 2 \, m_\p \, \pi \, k_B \, T_\mathrm{th} \right) \, ,
\end{equation}
and $T_\mathrm{th}$ is always greater than $(2 \, m_\p \, \pi \, k_B)^{-1}$. The
behavior of the temperature-dependent pressure in the model is depicted in Fig.
\ref{fig:p}.

\section{Simulation}

The simulation is done for a typical case of ball mill, that is spex mixer /
mill. However, the model under consideration can in principle be extended to
deal with another types of ball mills by changing the coordinate system
accordingly. 

\subsection{Coordinate system}

In the case of spex mill, the system contains a vial moving in a non-inertial
system \cite{delogu,concas}. Therefore, we can adopt the same coordinate system
and schematization of vial surface as Figs. 1 and 2 in \cite{concas}. Since the
system under consideration is moving on a non-inertial system $\vr = (x,y,z)$,
we should transform all coordinates in our formula to the coordinate system of
inertial system $(X,Y,Z)$. In the simulation, we make use of previous results on
the roto-translation transformation \cite{delogu,concas},
\begin{eqnarray}
 X(t) & = & x \, \cos\theta(t) \, \cos\alpha(t) 
	+ y \, \cos\theta(t) \, \sin\alpha(t) 
	\nonumber \\
	&& + z \, \sin\alpha(t) 
	+ L \, \sin\theta(t) \; ,\label{eq:xyz1}\\
Y(t) & = & - x \, \sin\alpha(t) + y \,\cos\alpha(t) \; , \label{eq:xyz2}\\
Z(t) & = & -x \, \sin\theta(t) \, \cos\alpha(t) 
	- y \, \sin\theta(t) \, \sin\alpha(t) 
	\nonumber \\
	&& + z \, \cos\alpha(t) 
	+ L \, \cos\theta(t) \; .
\label{eq:xyz3}
\end{eqnarray}
Here, $L$ is the length of mechanical shaft-arm, $\theta$ and $\alpha$ are the angles of rotation around the $Y-$ and $z-$axis. Both angles can be written as follows,
\begin{eqnarray}
 \theta & = & \theta_0 \, \sin(\omega \, t + \varphi) \; , \\
 \alpha & = & \alpha_0 \, \sin(\omega \, t + \varphi) \; ,
\end{eqnarray}
where $\theta_0$ and $\alpha_0$ are the angular momentum around the respected axis, $\omega$ is the frequency and $\varphi$ is the phase factor which depends on the initial conditions.

In our  simulation, for the sake of convenience the Cartesian coordinate is
transformed into the cylindrical coordinate system,
\begin{equation}
 (x,y,z) \longrightarrow (x, r \, \cos \vartheta, r \, \sin \vartheta) \; ,
\end{equation}
with $\vartheta$ is the rotation angle around $x$ in vial bases and $r \equiv \sqrt{y^2 + z^2}$.

As already mentioned above, we can apply the proposed model to any types of ball
mill. This can be accomplished by replacing the coordinate system of ball mill
under consideration like Eqs. (\ref{eq:xyz1})$\sim$(\ref{eq:xyz3}) to the
appropriate ones which represent its  geometrical motion.

\subsection{Technique}

\begin{figure}[t]
 \centering
 \includegraphics[width=6cm]{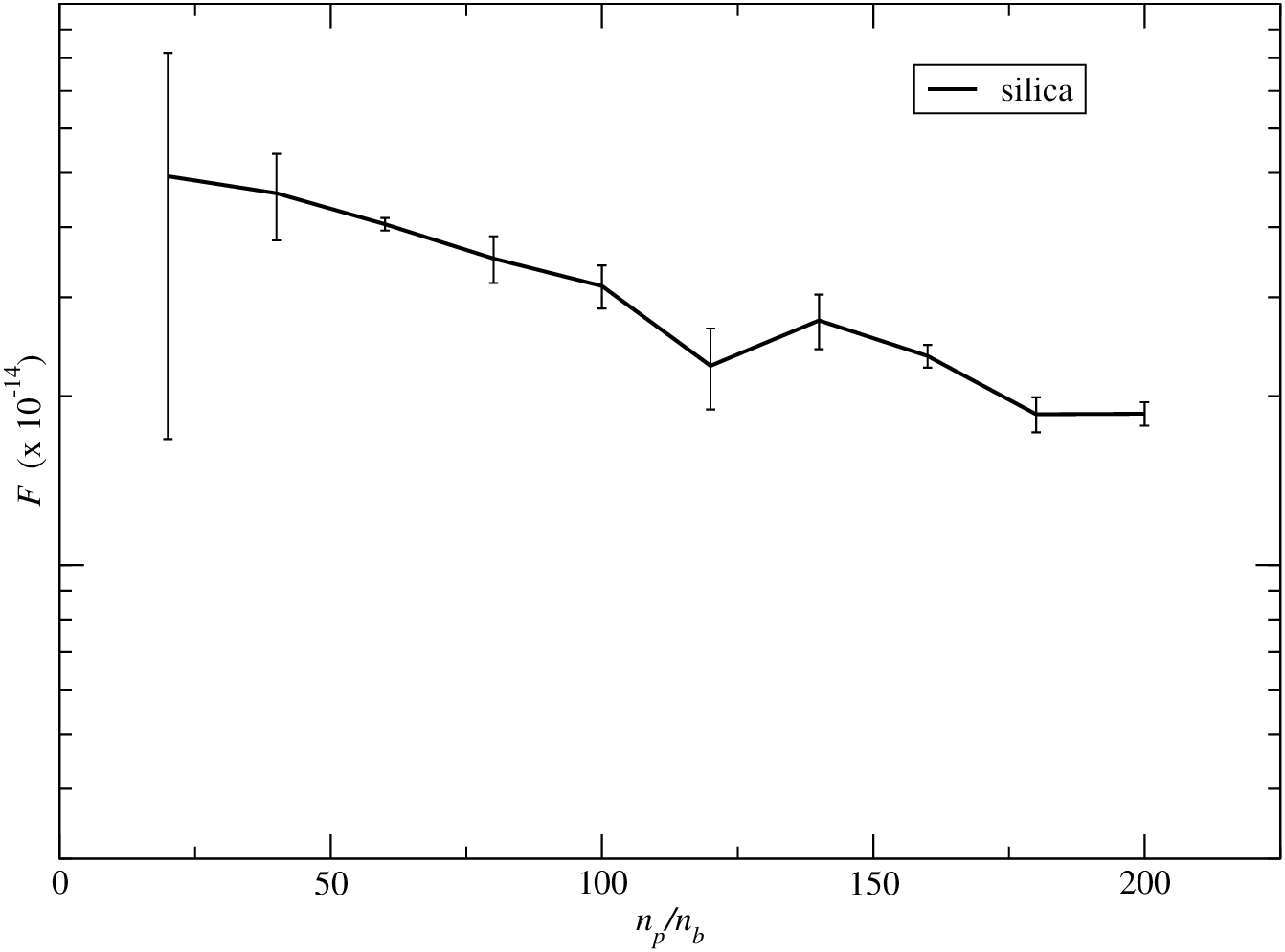} \hspace{5mm}
 \includegraphics[width=6cm]{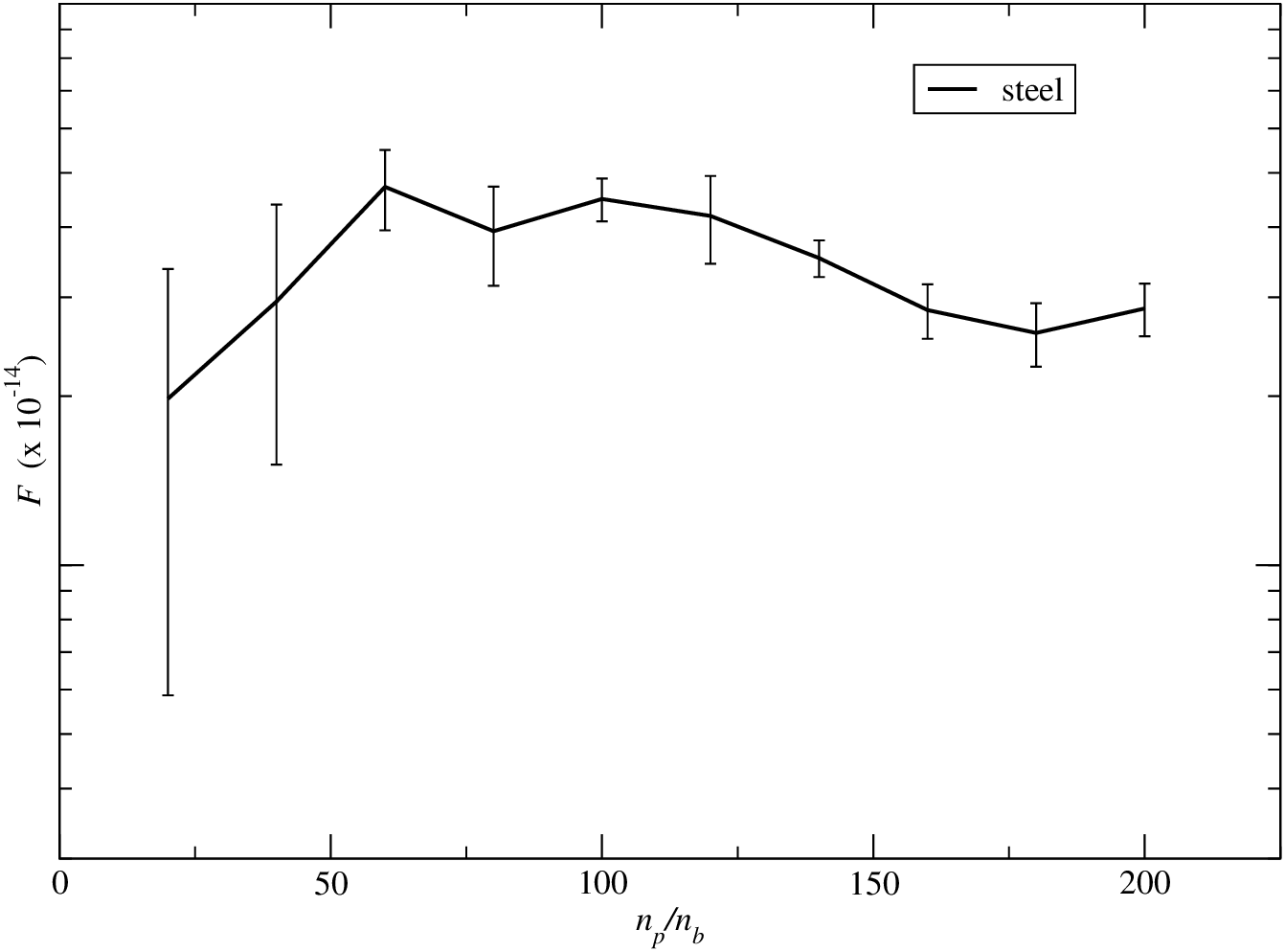} \\\vspace{1cm}
 \includegraphics[width=6cm]{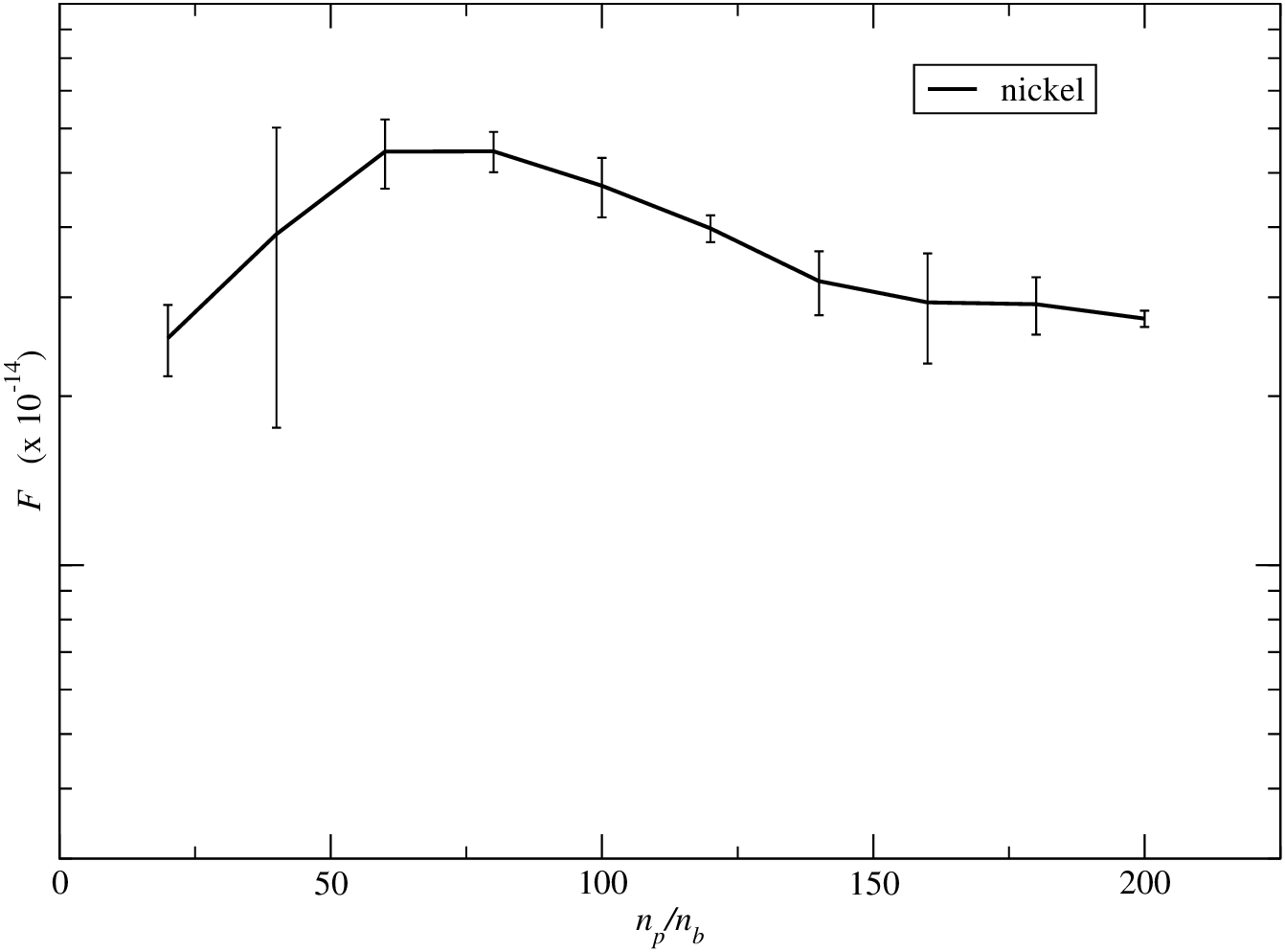}
 \caption{$\cf$ as a function of the ratio of matter number, ${n_\p}/{n_\b}$ for
silica, steel and nickel powders for ${R_\b}/{R_\p} = 50/3$.}
 \label{fig:n}
\end{figure}

The simulation within the present model and its underlying numerical
calculation are done using Monte Carlo technique. This is the most appropriate
technique to deal with higher dimensional integral
of many matters involved in the system like Eq. (\ref{eq:faux}) \cite{caflisch}.

We should note that the simulation here is performed to provide a complete
picture on the model and its applications rather than showing a comprehensive
numerical simulation that is out of the scope of present paper. A comprehensive
Monte Carlo integration in the present model requires proper 
resolution on 3-dimensional space which should be comparable with the
powder size, and also enough time resolution within full running period to
represent the whole dynamics. Unfortunately, this kind of simulation is very
time consuming, while the simulation itself is not the main interest in the
present paper. As an illustration, for $100 \mu$m powder size the 
appropriate resolution should be at least $\sim 10^4 \times 10^4 \times 10^4$.
The resolution must be increased accordingly as the powder size is
decreasing. Also, the present simulation is done for a particular time, that is
equivalent to a particular static position of vial motion. Note that in the full
simulation, the time evolution and the frequency of vial rotation are related
each other.

\begin{figure}[t]
 \centering
 \includegraphics[width=6cm]{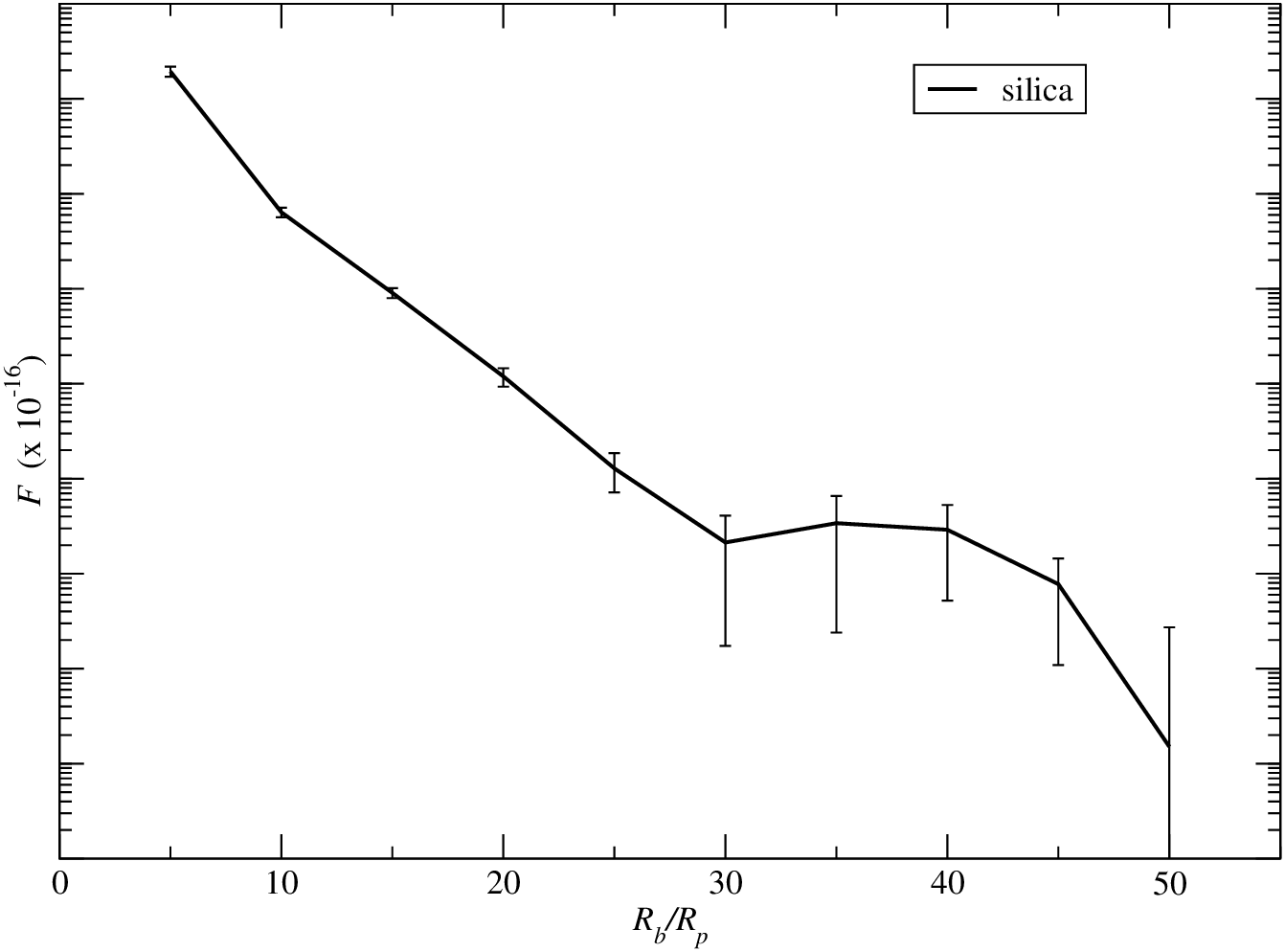} \hspace{5mm}
 \includegraphics[width=6cm]{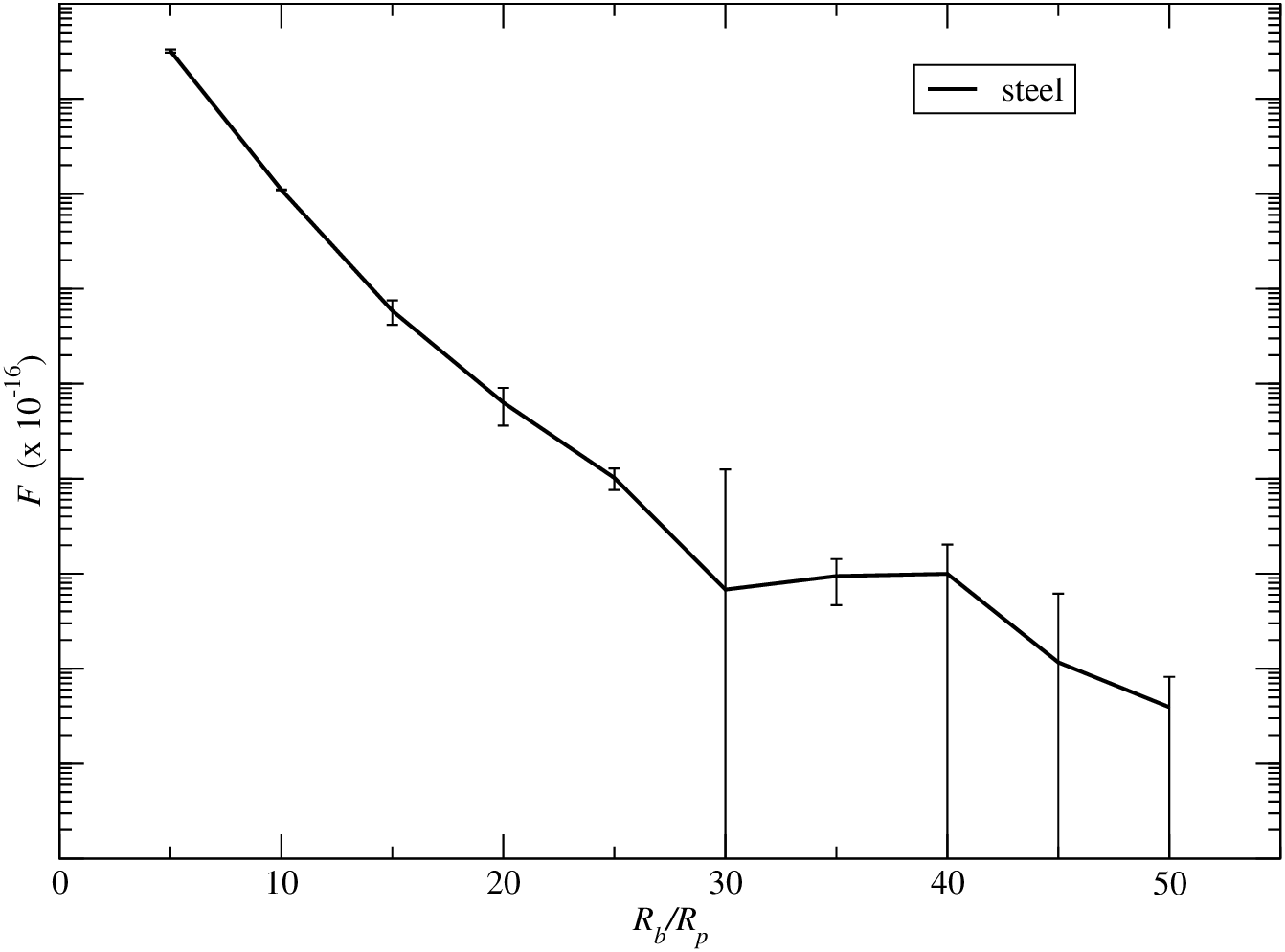} \\\vspace{1cm}
 \includegraphics[width=6cm]{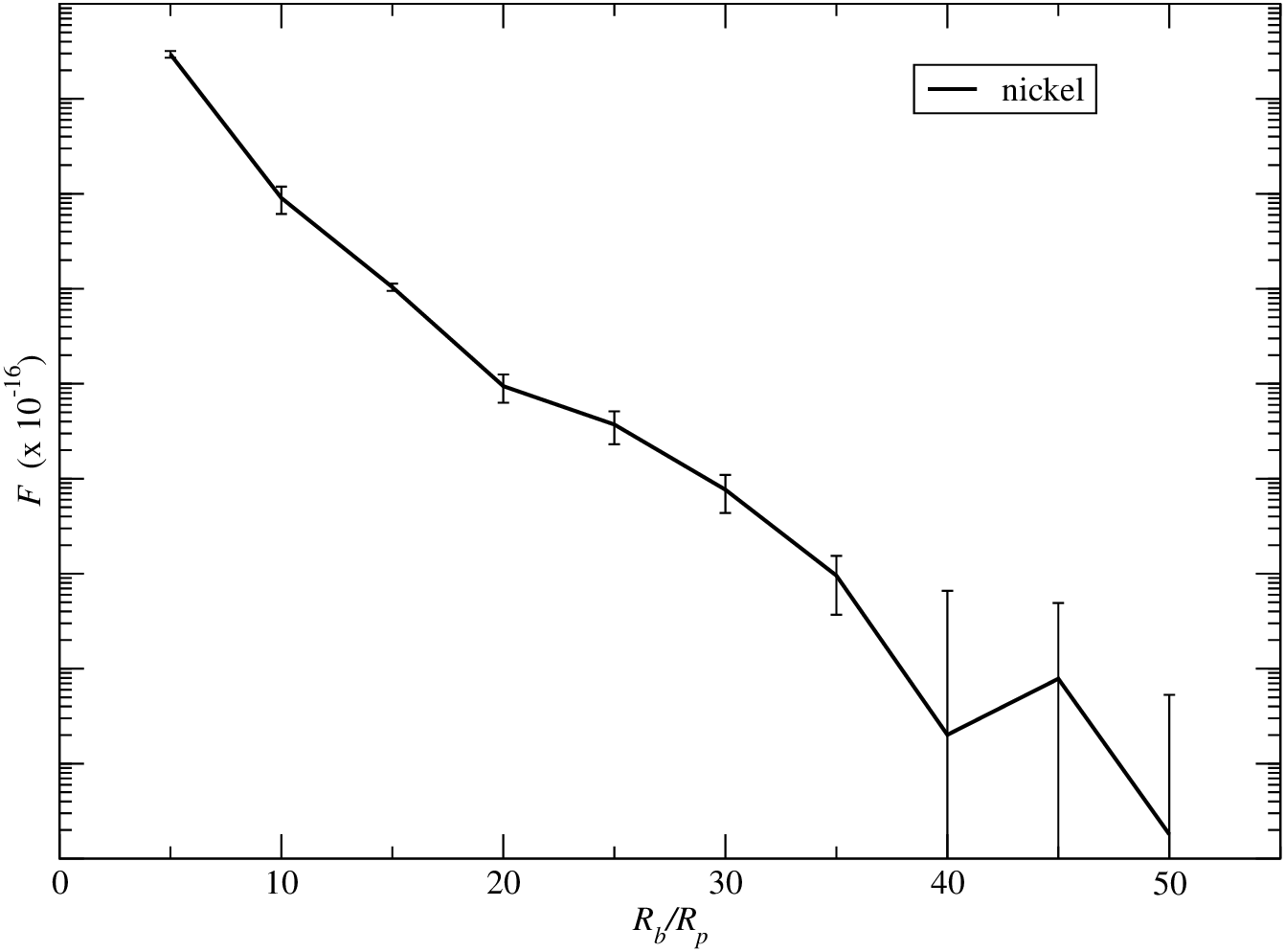}
 \caption{$\cf$ as a function of the ratio of matter size,
${R_\b}/{R_\p}$ for silica, steel and nickel powders for ${n_\p}/{n_\b} = 30$.}
 \label{fig:r}
\end{figure}

\subsection{Results}

Now we are ready to perform a preliminary simulation for the powders with
various strengths of interactions characterized by the defined parameters inside
the potentials, and also in a circumstances with non-zero electric field $\ve$
as well. The simulation is done for vial length $l = 50$ mm, vial radius $r_\v =
10$ mm, shaft-arm length $L = 200$ mm and ball radius $R_\b = 5$ mm.

The simulation of $\cf$ is performed for various values of the ratio of
matter number (${n_\p}/{n_\b}$) and matter size (${R_\b}/{R_\p}$). Each case is
also simulated for various materials of powders characterized with Young modulus
($\Upsilon$) and Poisson ratio ($v$) of
its sphere materials. The results are depicted in Figs. \ref{fig:n}
and \ref{fig:r}.

Figs. \ref{fig:n} and \ref{fig:r} show logarithmic scale of $\cf$ for silica,
steel and nickel powders. The error bars are coming from statistical errors due
to the uncertainties of Young modulus and Poisson ratio of each material. The
errors are significant for small ratio of matter number and large ratio of
matter size. These facts are natural since as large as the matter number ratio,
and also as small as matter size ratio would increase the probability of
collisions between balls and powders. On the other hand, the increasing matter
number and size ratios indicate the on-going comminution processes inside the
vial.

Again, it should be remarked that the present simulation has lost the
temperature evolution since it is done only for a single point of vial
position. The temperature evolution will be recovered in a full simulation
within the whole running period of ball mill.

\section{Summary}

We have proposed and discussed a novel model and approach for top-down
mechanical comminution processes to produce nanomaterial using ball mill
equipments. The study is focused on investigating the relevant
potentials in the hamiltonian for a ball mill system and the formalism to
extract relevant physical observables without tracking the geometrical
displacements inside the vial. The governing hamiltonian
is related to relevant physical observables through partition function of
statistical mechanics approach.

From  theoretical formalisms developed in this paper, we point out some
interesting remarks which hold for any type (and geometrical motions) of ball
mills,
\begin{itemize}
\item Any ball mills should have the same temperature dependencies of its 
normalized pressure as shown in Eq. (\ref{eq:pf}). Because the geometrical
structure is absorbed in the auxiliary function $\cf$ in Eq. (\ref{eq:faux}).
\item The magnetic field does not affect the matter dynamics inside the vial,
while the electric field does.
\item The contribution of dissipative term in the impact force is negligible
for small ratio of ${m_\p}/{m_\b}$ that is almost the case in all ball mills.
This fact simplifies the whole analysis
since the pre-knowledge of material viscosity is not needed anymore.
\item The model does not involve the powder number distribution, for
instance using the population balance algorithm etc \cite{bilgili}, in the
formalism. The static powder number is represented by the number of interacting
matters inside the vial, $n_\m$. Nevertheless, one can consider dynamic powder
number distribution using any breakage functions like $n_\m(t) \sim
n_{0_\m}(t_0) \, \mathrm{exp}(t/{t_0})$ with $n_{0_\m}$ is the initial powder
number.
\end{itemize}
According to these results, it can be argued that the model should easily be
confirmed by the experiments by verifying
the above features. More importantly, the experimental setup is much simpler
than the conventional approaches due to no requirement on tracking the
geometrical displacements of involving matters inside the vial.

We have also shown the simulation for particular case of spex mill using
Monte Carlo technique to calculate the complicated integrals in the formalism. 
More detail and comprehensive simulations are still under progress. Such works
involves some improvements. First of all, increasing the accuracy up to the
realistic powder size at the order of $\sim$ few tens nanometers. This means the
Monte
Carlo integration is done at the resolution of $10^8 \times 10^8 \times 10^8$.
Secondly, performing complete temperature (time) evolution with
appropriate accuracy within a realistic running period for various powder number
distributions. According to the present preliminary result, the proper time
resolution is $\sim O(\pi/\omega)$.

\section*{Acknowledgments}

The authors greatly appreciate inspiring discussion with N.T. Rochman and
A.S. Wismogroho throughout the work. FND thanks the Group for
Theoretical and Computational Physics LIPI for warm hospitality during the work.
This work is funded by the Riset Kompetitif LIPI in fiscal year
2010 under Contract no.  11.04/SK/KPPI/II/2010.

\bibliographystyle{elsarticle-num}
\bibliography{ballmill}

\end{document}